\documentstyle[epsfig]{elsart}
\newcommand{\lsim}{\mathrel{\rlap{\lower4pt\hbox{\hskip1pt$\sim$}}
\raise1pt\hbox{$<$}}}
\newcommand{\sfrac}[2]{\mbox{\footnotesize $\frac{#1}{#2}$}}
\begin{document}
\begin{frontmatter}
%_______________________ Title, Authors ____________________________________
{\hspace*{\fill}{\sc Preprint Numbers}: \parbox[t]{45mm}
{ANL-PHY-8585-TH-96\\ TRI-PP-96-63}}
% nucl-th/yymmddd

\title{$K_{\ell 3}$ and $\pi_{e 3}$ transition form factors}
\author[dubna]{Yu. Kalinovsky,}
\author[triumf]{K. L. Mitchell}
\author[anl]{and C. D. Roberts}
\address[dubna]{Bogoliubov Laboratory of Theoretical Physics, \\
Joint Institute for Nuclear Research, 141980 Dubna, Russia}
\address[triumf]{TRIUMF, 4004 Wesbrook Mall,\\
Vancouver, British Columbia, Canada V6T 2A3}
\address[anl]{Physics Division, Bldg. 203, Argonne National Laboratory,\\
Argonne IL 60439-4843, USA}
%-------------------------------------------------------------------
\begin{abstract}
$K_{\ell 3}$ and $\pi_{e 3}$ transition form factors are calculated as an
application of Dyson-Schwinger equations.  The role of nonanalytic
contributions to the quark--W-boson vertex is elucidated.  A one-parameter
model for this vertex provides a uniformly good description of these
transitions, including the value of the scalar form factor of the kaon at the
Callan-Treiman point.  The $K_{\ell 3}$ form factors, $f_\pm^K$, are
approximately linear on $t\in [m_e^2,m_\mu^2]$ and have approximately the
same slope.  $f_-^K(0)$ is a measure of the Euclidean constituent-quark mass
ratio: $M^E_s/M^E_u$.  In the isospin symmetric limit: $-f_+^\pi(0)=
F_\pi(t)$, the electromagnetic pion form factor, and $f_-^\pi(t)\equiv 0$.
\end{abstract}
%\pacs{13.20.-v, 14.40.Aq, 12.38.Lg, 24.85.+p}
\end{frontmatter}
%____________________________________________________________

{\bf 1. Introduction}.\\ The semileptonic transitions $K^+\to \pi^0 \ell
\nu_\ell$ [$K_{\ell 3}^+$], $K^0\to \pi^- \ell \nu_\ell$ [$K_{\ell 3}^0$] and
$\pi^+\to \pi^0 e \nu_e$ [$\pi_{e 3}$] proceed via the flavour-changing,
vector piece of the $V-A$ electroweak interaction, in particular $j_\mu^{su}$
and $j_\mu^{du}$.  The axial-vector component does not contribute because, in
every case, the two mesons involved have the same parity.  Neither
$j_\mu^{su}$ nor $j_\mu^{du}$ is conserved; in each case the symmetry
breaking term is a measure of the current-quark mass difference, $m_s-m_u$ or
$m_d-m_u$, and its enhancement due to nonperturbative effects.  Therefore
these processes can be employed to probe $SU_f(3)$ flavour symmetry
violation.

The current status of experimental analyses of $K_{\ell 3}^+$ and $K_{\ell
3}^0$ transitions is summarised in Ref.~\cite{PDG94}.  It is not completely
satisfactory, with the value of two of the four observables showing a
surprising sensitivity to the initial state (see Table~\ref{tableA}).
Contemporary theoretical analyses fall into two classes: those employing
quark-gluon degrees of freedom; e.g., Refs.~\cite{SI92,AB96}, and those using
meson degrees of freedom; e.g., Refs.~\cite{LR84,GL85}.  While there is
agreement between these approaches for some observables, in Sect.~3 we
identify {\it qualitatively important} quantitative disagreements whose
resolution requires an appreciation of the origin and role of nonanalytic
contributions to the quark-W-boson vertex.

Our analysis of the transition amplitudes, and their form factors, is a
phenomenological application of Dyson-Schwinger equations [DSEs].  It is an
extension of the calculation of the electromagnetic form factors:
$F_{K^\pm}(t)$, $F_{K^0}(t)$~\cite{BRT96} and $F_\pi(t)$~\cite{R96}.  As
therein, primary elements of this calculation are the dressed-quark
propagator (2-point Schwinger function) for the $u$-, $d$- and $s$-quarks and
the Bethe-Salpeter amplitudes for the $\pi$- and $K$-mesons, the behaviour of
which follows from extensive nonperturbative, model-DSE
studies~\cite{DSErev}.  

A significant new feature of this study is the dressed-vertex (3-point
Schwinger function) describing the quark-$W$-boson coupling.  A qualitative
understanding of this is crucial in resolving the discrepancies between
quark- and meson-based analyses of the transition form factors; and is an
important precursor to the study of other weak-interaction processes such as
those involving baryons.

{\bf 2. Transition Form Factors.}\\ The matrix elements for the $K_{\ell 3}$
and $\pi_{e 3}$ transitions are
\begin{eqnarray}
\label{jkplus}
J_\mu^{K^+}(K,Q) \equiv
\langle \pi^0(p) | \bar u\gamma_\mu s | K^+(k)\rangle & \equiv &
\sfrac{1}{\surd 2}\,\left(f_+^{K^+}(t)\, K_\mu 
+ f_-^{K^+}(t) \,Q_\mu \right)\\
\label{jzero}
J_\mu^{K^0}(K,Q) \equiv
\langle \pi^-(p) | \bar u\gamma_\mu s | K^0(k)\rangle & \equiv &
f_+^{K^0}(t) \, K_\mu + f_-^{K^0}(t) \,Q_\mu \\
\label{jpiplus}
J_\mu^{\pi}(K,Q) \equiv
\langle \pi^0(p) | \bar u\gamma_\mu d | \pi^+(k)\rangle & \equiv &
\sfrac{1}{\surd 2}\,\left(f_+^{\pi}(t)\, K_\mu 
+ f_-^{\pi}(t) \,Q_\mu \right)\,,
\end{eqnarray}
where $K= k+p$, $Q=k-p$ and the squared-momentum transfer $t=-Q^2$.
\footnote{We employ a Euclidean space formulation with
$\{\gamma_\mu,\gamma_\nu\}=2\delta_{\mu\nu}$, $\gamma_\mu^\dagger =
\gamma_\mu$ and $a\cdot b=\sum_{i=1}^4 a_i b_i$.  A timelike vector, $Q_\mu$,
has $Q^2<0$.}

In the isospin-symmetric case, $m_u=m_d$, $j_\mu^{du}$ is conserved and: 1)
$f_\pm^{K^0}\equiv f_\pm^{K^+}$; 2) $f_+^{\pi}(t)=-F_\pi(t)$; and 3)
$f_-^{\pi}\equiv 0$, while in the case of $SU_f(3)$ symmetry, $j_\mu^{su}$ is
also conserved, $f_+^{K^0,K^+,\pi}(t)\equiv -F_\pi(t)$ and
$f_-^{K^0,K^+,\pi}\equiv 0$.  Away from the symmetry limits, the
Ademollo-Gatto theorem~\cite{AGT} entails $f_+^2(0) \approx 1$; i.e., in the
vector form factor, $SU_f(3)$ symmetry breaking effects are suppressed at
$t=0$.  However, one expects $f_-(0)$ to be sensitive to the nonperturbative
enhancement of the $SU_f(3)$ symmetry breaking current-quark mass
differences, since $f_-(0)$ depends on the $s$:$u$ ratio of constituent-quark
masses in Ref.~\cite{NI75}.

The scalar form factor
\begin{equation}
f_0^K(t) \equiv f_+^K(t) + \frac{t}{m_K^2 - m_\pi^2} f_-^K(t)
\end{equation}
measures the divergence of the currents in Eqs.~(\ref{jkplus}) and
(\ref{jzero}), which makes it a particularly useful observable.  Current
algebra predicts the value of $f_0^K$ at the Callan-Treiman point,
$t=m_K^2-m_\pi^2\equiv \Delta$: $f_0^K(\Delta)= -f_K/f_\pi$, the ratio of the
weak-decay constants~\cite{CT66}.  The Callan-Treiman point is not accessible
experimentally, however, the robust nature of the derivation of this result
makes it a useful tool in constraining and improving a given theoretical
framework.

In impulse approximation
\begin{eqnarray}
\label{tff}
\lefteqn{J_\mu^{K^+}(K,Q)= \surd 2 N_c \int\frac{d^4 \ell}{(2\pi)^4}
{\rm tr}_D\left[
\bar\Gamma_{\pi^0}\left(\ell + \sfrac{1}{4} [K + Q];-p\right)\times
\right.
}\\
&& \nonumber
\left.
S_u(\ell + \sfrac{1}{2} K)
\Gamma_{K^+}\left(\ell + \sfrac{1}{4}[K - Q];k\right)
S_s(\ell - \sfrac{1}{2} Q)
iV_\mu^{su}(\ell;-Q)
S_u(\ell+\sfrac{1}{2} Q)\right]\,,
\end{eqnarray}
with obvious modifications for $K_{\ell 3}^0$ and $\pi_{e 3}$.  In
Eq.~(\ref{tff}): $S_f$ is the dressed-propagator for quark flavour $f$;
$\Gamma_M(q;P)$ is the Bethe-Salpeter amplitude for a meson $M$, with $q$ the
relative quark-antiquark momentum and $P$ the total momentum; and
$V_\mu^{su}(q;P)$ is the dressed $s$-$u$-$W$-vertex.  Herein we work in the
isospin symmetric limit; i.e., we do not distinguish between $u$- and
$d$-quarks, and hence $S_u\equiv S_d$.

Extensive studies of the DSE for the dressed-quark propagator~\cite{DSErev}
lead to the following algebraic approximating form, used successfully in
Refs.~\cite{BRT96,PL96}: $S_f(p) = -i\gamma\cdot p \sigma_V^f(p) +
\sigma_S^f(p)$,
\begin{eqnarray}
\label{SSM}
\bar\sigma^f_S(x)  & =  & C^f({\bar m_f})\, {\rm e}^{-2x}  + 
        \frac{\bar m_f}{x + \bar m_f^2}
                \left( 1 - e^{- 2\,(x + \bar m_f^2)} \right)\\
& & \nonumber
      + \frac{1 - e^{- b^f_1 x}}{b^f_1 x}\,\frac{1 - e^{- b^f_3 x}}{b^f_3 x}\,
        \left( b^f_0 + b^f_2 \frac{1 - e^{- \Lambda x}}{\Lambda\,x}\right) \\
\label{SVM}
\bar\sigma^f_V(x) & = & \frac{2 (x+\bar m_f^2) -1 
                + e^{-2 (x+\bar m_f^2)}}{2 (x+\bar m_f^2)^2}
                - \bar m_f C^f({\bar m_f})\, e^{-2 x},
\end{eqnarray}
where $x=p^2/(2 D)$ and: $\bar\sigma_V^f(x) = 2 D\,\sigma_V^f(p^2)$;
$\bar\sigma_S^f(x) = \sqrt{2 D}\,\sigma_S^f(p^2)$; and $\bar m_f$ =
$m_f/\sqrt{2 D}$, with $D$ a mass scale.  We write the inverse of the
propagator as $S_f^{-1}(p) = i\gamma\cdot p\, A^f(p^2) + B^f(p^2)\,$.

The Bethe-Salpeter equation [BSE] for the pseudoscalar mesons has also been
studied extensively~\cite{DSErev}.  The realisation of Goldstone's theorem
entails that, in the chiral limit, $m_f=0$, the pseudoscalar Bethe-Salpeter
amplitude is completely determined by the dressed-quark
propagator~\cite{BRS96}.  As a consequence one finds that
\begin{eqnarray}
\label{pibsa}
\Gamma_\pi(p;P^2=-m_\pi^2)  &\approx  &
        i\gamma_5\,\frac{1}{f_\pi}\, B_{m_u=0}^u(p^2)\,, \\
\label{kbsa}
\Gamma_K(p;P^2=-m_K^2)  &\approx  &
        i\gamma_5\,\frac{1}{f_K}\, B_{m_s=0}^s(p^2)~,
\end{eqnarray}
with $f_\pi\approx 92\,$MeV and $f_K\approx 113\,$MeV the calculated pion and
kaon normalisation constants, respectively, are good pointwise
approximations.  These results follow from BSE studies and have proven
phenomenologically efficacious~\cite{BRT96,PL96}.

The parameters $C^f({\bar m_f})$, $\bar m_f$, $b_{1\ldots 3}^f$ in
Eqs.~(\protect\ref{SSM}), (\protect\ref{SVM}) are determined in a
$\chi^2$-fit to a range of hadronic observables.  This is described in
Ref.~\cite{BRT96} and leads to the values in Table~\ref{tableB}.  $C^f({\bar
m_f})$ is a function of $\bar m_f$; it is only non-zero when evaluating the
Bethe-Salpeter amplitudes using Eqs.~(\protect\ref{pibsa}),
(\protect\ref{kbsa}), which allows the algebraic approximation to represent
those differences between $B_{m=0}(0)$ and $B_{m\neq 0}(0)$ observed in
numerical studies of the quark DSE.  In the fit, the difference between the
$u$- and $d$-quarks was neglected and only that minimal difference between
$u$- and $s$-quarks allowed that was necessary to ensure: $\langle \bar s
s\rangle <\langle \bar u u\rangle$; and $m_s/m_u\gg 1$.  ($\Lambda=10^{-4}$
is included in Eqs.~(\protect\ref{SSM}), (\protect\ref{SVM}) only for the
purpose of separating the small- and intermediate-$p^2$ behaviour of the
algebraic form, characterised by $b_0$ and $b_2$; a separation in magnitude
observed in numerical studies.)

\begin{table}[h,t]
\begin{center}
\caption{\label{tableB} The values of $b_{1,3}^s$ are underlined
to indicate that the constraints $b_{1,3}^s=b_{1,3}^u$ were imposed in the
fitting.  The scale parameter $D=0.160\,$GeV$^2$.}

\begin{tabular}{ccccccc}\hline
    & $C^f({\bar m_f=0})$ & $\bar m_f$ & $b_0^f$ & $b_1^f$ & $b_2^f$& $b_3^f$ 
\\\hline
$u$ & 0.121 & 0.00897 & 0.131 & 2.90 & 0.603 & 0.185 \\
$s$ & 1.69  & 0.224   & 0.105 & \underline{2.90} & 0.740 & \underline{0.185}
\\\hline
\end{tabular}
\end{center}
\end{table}

The algebraic approximation represented by Eqs.~(\ref{SSM})-(\ref{kbsa})
provides a good description of pion and kaon observables~\cite{BRT96} and has
recently been employed successfully in the prediction of a wide range of
vector-meson electroproduction observables~\cite{PL96}; a process far outside
the domain on which it was constrained.

{\bf 2.1}~{\it Dressed Quark W-Boson Vertex}.  $V_\mu^{su}(p;P)$ in
Eq.~(\ref{tff}) is the new element in this study.  It satisfies a DSE, from
which one can derive the following Ward-Takahashi identity
\begin{equation}
Q_\mu\,iV_\mu^{su}(p;Q) = S^{-1}_s(p_+) - S^{-1}_u(p_-)
- (m_s - m_u)\,\Gamma_I^{su}(p;Q)\,
\label{WTI}
\end{equation}
where $p_\pm \equiv p\pm Q/2$ and $\Gamma_I^{su}(p;Q)$ is the
flavour-dependent scalar vertex (neglecting interactions, $\Gamma_I^{su}(p;Q)
= I_D$).  This entails that the flavour-changing vector current is not
conserved and that the symmetry breaking term measures the current-quark mass
difference.  Nonperturbative contributions to $\Gamma_I^{su}(p;Q)$ enhance
this effect.

Herein we use what is known from extensive studies of the flavour-dependent,
electromagnetic quark-photon vertex, $\Gamma_\mu^f(p;Q)$, to construct an
Ansatz for $V_\mu^{su}(p;P)$.  In studies of pion and kaon electromagnetic
form factors~\cite{BRT96,R96} and vector meson electroproduction~\cite{PL96}
the Ball-Chiu Ansatz~\cite{BC80}
\begin{equation}
i\Gamma_\mu^f(p;Q) = i \gamma_\mu \,f_1^f(p;Q) 
        + i\gamma\cdot p\, p_\mu \,f_2^f(p;Q)
        + p_\mu \,f_3^f(p;Q)\,,
\end{equation}
where 
\begin{equation}
\begin{array}{l}
f_1^f(p;Q)= [A^f(p_+)+A^f(p_-)]/2,\\
f_2^f(p;Q)=[A^f(p_+)-A^f(p_-)]/p\cdot Q,\\
f_3^f(p;Q)= [B^f(p_+)-B^f(p_-)]/p\cdot Q
\end{array}
\end{equation}
has proven successful.  As described in Ref.~\cite{R96}, this Ansatz is
successful because it is the minimal solution of the Abelian Ward-Identity
that is completely determined by the dressed-quark propagator and: A) has a
well defined limit as $Q^2\to 0$; B) transforms correctly under $C$, $P$, $T$
and Lorentz transformations; and C) reduces to the bare vertex in the manner
prescribed by perturbation theory.  It therefore satisfies four of the six
physical constraints proposed in Ref.~\cite{BR93} and explored in detail in
Ref.~\cite{BP94}. Consequently, a first, minimal Ansatz is
\begin{equation}
\label{ansatza}
V_\mu^{su}(p;Q) = \frac{1}{2}\left(\Gamma_\mu^s(p;Q) 
                + \Gamma_\mu^u(p;Q)\right)\,,
\end{equation}
which satisfies Eq.~(\ref{WTI}), with a particular form of the symmetry
breaking term, and also A)-C) above.  

A property that our study, when using Eq.~(\ref{ansatza}), shares with other
quark-based studies; for example, Refs.~\cite{SI92,AB96,NI75}, is that the
$s$-$u$-$W$-vertex is analytic on the real-$t$ axis.  An Ansatz with this
property excludes resonance contributions, such as $K$-$\pi$ loops, which
although unimportant for $t<0$, may provide significant contributions for
$t\in[m_e^2,m_\mu^2]$.  We now address this issue.

{\bf 3. Results and Discussion}.\\
Equations~(\ref{tff})-(\ref{kbsa}) and (\ref{ansatza}) yield the results in
the first column of Table~\ref{tableA}.  They are in quantitative agreement
with the results of Refs.~\cite{SI92,AB96,NI75} and, with some {\it
qualitatively important} exceptions that we discuss below, with
Refs.~\cite{LR84,GL85}.

Before addressing these exceptions we discuss other points of interest.  The
result for $f_+(0)$ is consistent with the Ademollo-Gatto theorem.  Further,
the calculated value of $f_-(t_m)$ can be compared with Ref.~\cite{NI75}.
Using our dressed-quark propagators we define a Euclidean constituent-quark
mass, $M^E_f$,~\cite{PW91} which provides a single, indicative and
quantitative measure of the importance of nonperturbative dressing of the
quarks in the infrared: $(M^E_f)^2$ is the solution of $p^2 [A^f(p^2)]^2 -
[B^f(p^2)]^2=0$.  In the present case we find $M^E_{u,d}=330\,$MeV,
$M^E_s=490\,$MeV.  From Fig.~1 of Ref.~\cite{NI75} these values lead to
$f_-(t_m)\sim 0.28$, consistent with our result.  (There is a relative
``$-\,$''sign between our definitions of $f_\pm$ and those of
Ref.~\cite{NI75}.) This emphasises that $f_-(0)$ is a probe of the
nonperturbative enhancement of $SU_f(3)$ breaking.

In the experimental analysis of $K_{\ell 3}$ decays it is often assumed that,
on $t\in [m_e^2,m_\mu^2]$, $f_+(t)$ is linear; i.e.,
$f_+(t)=f_+(0)[1+\lambda_+ t/m_\pi^2]$, and $f_-(t)=f_-(0)[1+\lambda_-
t/m_\pi^2]$, with $\lambda_-=0$; i.e., $f_-(t)=\,$constant.  Using the
definition of $\lambda$ in Table~\ref{tableA} we find that the difference
$\lambda_+^e-\lambda_+^\mu$ is small and hence that a linear approximation to
$f_+$ is reasonable.  However, in our study we find $\lambda_- \sim
\lambda_+$, consistent with Refs.\cite{SI92,AB96} where
$\lambda_+=\lambda_-$, but inconsistent with the assumption that
$f_-(t)=\,$constant.

\begin{table}[h,t]
\begin{center}
\caption{$t_m = (m_K-m_\pi)^2$ is the largest, physically accessible value of
the squared momentum-transfer, $t$; $\xi(t)\equiv f_-(t)/f_+(t)$;
$\lambda_\alpha^\ell\equiv m_{\pi^+}^2
f_\alpha^\prime(m_\ell^2)/f_\alpha(0)$, $\alpha\in\{+,-,0\}$; and $r_{\pi
K}^2 \equiv 6 f_+^\prime(0)/f_+(0)$.  Experimental results are taken from
Ref.~\protect\cite{PDG94}.  $m_e^2/m_\mu^2\ll 1$ means that $\lambda_0^e$ is
not accessible experimentally.  References to theoretical comparisons are
labelled: $a=$ Ref.~\protect\cite{SI92}, $b=$ Ref.~\protect\cite{AB96}, $c=$
Ref.~\protect\cite{LR84}, $d=$ Ref.~\protect\cite{GL85}; where necessary,
results have been calculated using the information provided.
\label{tableA}}
\begin{tabular}{cccccc}\hline
                & Eq.~(\protect\ref{ansatza}) &Eq.~(\protect\ref{ansatzb}) 
                & Comparison & Experiment: 
                                $\begin{array}{c}
                                K_{\mu3}^+\\
                                K_{\mu3}^0\end{array}$   \\\hline
$-f_+(t_m)$      & $1.11$      &  $1.24$                    
                       & $1.04^a$ &  \\
$f_-(t_m)$      & $0.27$      &    $0.30$       
                        & $0.29^a$ &  \\
$-f_+(0)$        & $0.98$     &    $0.98$       
                        &    $0.93^a$, $0.98^c$    &  \\ 
$f_-(0)$        & $0.24$      &    $0.24$       & $0.26^a$, $0.15^d$
                                                & \\
$-\xi(0)$        & $0.25$      &   $0.25$                    
                                        & $0.28^a$,
                                          $0.28^b$,
                                          $0.15^d$
                                              &$
                                        \begin{array}{c}
                                        0.35 \pm 0.15 \\
                                        0.11 \pm 0.09
                                        \end{array}$\\ 
$\lambda_+^e$   & $0.018$     &    $0.030$                   
                                        & $0.019^a$,
                                         $0.028^b$,
                                         $0.030^d$ 
                                        &$\begin{array}{c}
                                         0.0286\pm 0.0022\\
                                         0.0300\pm 0.0016
                                        \end{array}$\\ 
$\lambda_+^\mu$ & $0.018$     &     $0.031$                   
                                        & $0.019^a$,
                                         $0.028^b$,
                                        $0.030^d$
                                        &$\begin{array}{c}
                                        0.033\pm 0.008\\  
                                        0.034\pm 0.005
                                        \end{array}$\\ 
$\lambda_-^e$   & $0.012$     &     $0.024$                  
                                        & $0.019^a$,
                                         $0.028^b$,
                                                && \\
$\lambda_-^\mu$ & $0.012$     &      $0.025$                 
                                        & $0.019^a$,
                                         $0.028^b$,
                                                && \\
$-f_0(\Delta)$   & $0.95$     &      $1.22$                 
                                        & $0.88^a$,
                                          $1.22^d$
                                                && \\
$\lambda_0^e$   & $-0.0024$   &      $0.0082$                 
	                                       & $-0.005^a$,
                                          $0.0026^b$,
                                          $0.017^d$
                                                & \\
$\lambda_0^\mu$ & $-0.0024$   &       $0.0089$                 
	                                       & $-0.005^a$,
                                          $0.0026^b$,
                                        $ 0.017^d$
                                        &$\begin{array}{c}
                                        0.004\pm 0.007\\
                                        0.025\pm 0.006
                                        \end{array}$\\ 
$r_{\pi K}\,$(fm)
                & $0.47$      &         $0.60$              
                                        & $0.48^a$, 
                                        $ 0.60^d$
                                        & \\\hline
\end{tabular}
\end{center}
\end{table}

In addressing the exceptions, we note from Table~\ref{tableA} that, in
contrast to Ref.~\cite{GL85}, neither our calculation using
Eq.~(\ref{ansatza}) nor Refs.\cite{SI92,AB96} reproduce the anticipated value
of the scalar form factor at the Callan-Treiman point, $f_0(\Delta)=
-f_K/f_\pi$. Correlated with this are the results for $\lambda_0$: in
comparison with the experimental analysis of $K_{\mu 3}^0$ decays, which have
the smallest error, $\lambda_0$ is of the wrong-sign or, if of the correct
sign, it is an order-of-magnitude too small.  These results are systematic
effects due to the magnitude of $f_\pm(0)$ and the fact that $\lambda_+
\simeq \lambda_-$.  To establish this we employed many variants of
Eq.~(\ref{ansatza}), the simplest of which replaces $[\Gamma_\mu^s +
\Gamma_\mu^u]/2$ by $\alpha\Gamma_\mu^s + (1-\alpha)\Gamma_\mu^u$,
$0<\alpha<1$.  The various Ans\"atze could introduce quantitative changes of
$\sim$ 5\% in $f_0(\Delta)$ and $\sim$ 50\% in $\lambda_0$, however, such
changes are patently inadequate.

{\bf 3.1}~{\it Nonanalytic Contributions to the Vertex}.  This discrepancy is
a defect of the vertex Ansatz, Eq.~(\ref{ansatza}); a defect that is implicit
in Refs.~\cite{SI92,AB96} and similar studies.  That this is the case is
signalled by the small value of $r_{\pi K}$ in the first column of
Table~\ref{tableA}.  Nonanalytic, resonance contributions to the quark-photon
vertex or, analogously, $\pi$-$\pi$ rescattering, provide $\lsim$ 10\% of
$r_\pi\,$\cite{ABR95}, which measures $F_\pi^\prime(t= 0)$.  Such nonanalytic
contributions rapidly becomes insignificant as $(-t)$ increases because
resonances do not contribute to the vertex at spacelike momentum
transfer~\cite{MT96}.  However, their importance increases in the timelike
region and, in calculating $f_0(t=\Delta\approx 0.56 (m_K +m_\pi)^2)$, one is
nearing the $K$-$\pi$ production threshold where such contributions dominate.

In a careful study of the DSE for the $s$-$u$-$W$-vertex, these contributions
are manifest in the solution at $t>0$.  They should therefore be included in
constructing a realistic Ansatz for $V_{\mu}^{su}$ and hence, following
Ref.~\cite{ABR95}, we consider a heuristic minimal modification of
Eq.~(\ref{ansatza}), redefining $f_1$ as follows:
\begin{equation}
\label{ansatzb}
f_1(k;Q)   \to  f_1(k;Q) {\cal M}(t), \; 
{\cal M}(t) \equiv
\left[1
+ c_l\,{\rm e}^{t/t_p}\,{\cal L}(t)\right]\,,
\end{equation}
with $t_p= (m_K+m_\pi)^2$, and where
\begin{eqnarray}
{\cal L}( t)  &\equiv  &
2 + \ln\left[\frac{m_\pi^2}{m_K^2}\right] 
        \left( \frac{\Delta}{t} - \frac{\Sigma}{\Delta}\right)
        - \frac{\nu(t)}{t}
        \ln\left[\frac{(t+\nu(t))^2-\Delta^2}
                {(t-\nu(t))^2-\Delta^2}\right],
\end{eqnarray}
with $\Sigma=m_K^2+m_\pi^2$ and $\nu(t)^2=(t-t_p)(t-t_m)$, expresses the
essence of the nonanalytic structure of the $K$-$\pi$ loop~\cite{GL85}.  In
Eq.~(\ref{ansatzb}), $c_l$ parametrises the relative weight of the analytic
and nonanalytic terms at $t=\Delta$.  Its value is determined by the
requirement that this Ansatz yield $f_0(\Delta)= -f_K/f_\pi$.  This Ansatz,
which herein is only sampled on the domain $t\in(-\infty,\Delta]$, is only a
useful heuristic tool if the required value of $c_l$ is small, for then it
satisfies our physical requirements: it leaves the vertex unmodified for
spacelike-$t$; and it has a logarithmic branch point at the $K$-$\pi$
production threshold.

The results obtained using Eq.~(\ref{ansatzb}), with $c_l=0.17$, are
presented in the second column of Table~\ref{tableA}.  With this value of
$c_l$, $0.94<{\cal M}(t<0)\leq 1$; i.e., for $t<0$, the modification of the
vertex is small.  The value of the slope parameters, $\lambda_{+,0}$, are in
agreement with experiment and Ref.~\cite{GL85}, as is $r_{\pi K}$.  As
expected, $f_\pm(0)$ and $\xi(0)$ are unchanged.  This outcome is
qualitatively insensitive to the exact form of ${\cal L}$; for example, using
${\cal L}(t)= - \{1 + (t_p/t) \ln[1-t/t_p]\}$ with $c_l=0.335$, leads only to
a $\sim 3$\% reduction in $f_\pm(t_m)$ and a $\sim 27$\% reduction in
$\lambda_0$.

\begin{figure}[t,h]
\centering{\
\epsfig{figure=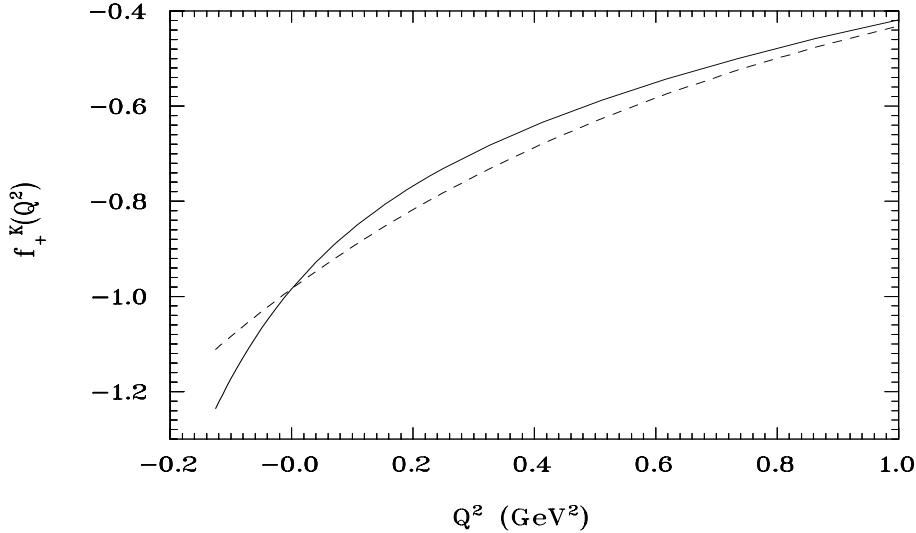,height=7cm,rheight=7cm,width=12cm,angle=90  }}
\caption{$f^K_+(Q^2)$: solid line, Eq.~(\protect\ref{ansatzb}); dashed line,
Eq.~(\protect\ref{ansatza}).  $f_+^K$ is approximately linear for
$-t_m<Q^2<0$.
\label{figa}}
\end{figure}
\begin{figure}[h,t]
\centering{\
\epsfig{figure=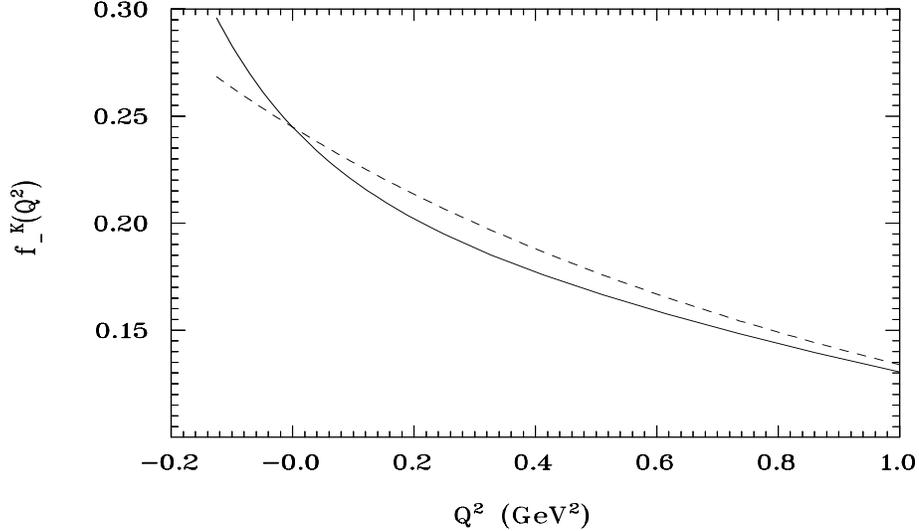,height=7cm,rheight=7cm,width=12cm,angle=90  }}
\caption{$f^K_-(Q^2)$: solid line, Eq.~(\protect\ref{ansatzb}); dashed line,
Eq.~(\protect\ref{ansatza}).  $f_-^K$ is approximately linear for
$-t_m<Q^2<0$ and it varies almost as rapidly as $f_+^K$.
\label{figc}}
\end{figure}
In Figs.~\ref{figa} and \ref{figc}, to further illustrate the effect of
nonanalytic contributions, we plot the $K_{\ell 3}$ transition form factors.
$K_{\ell 3}^+$ is not distinguished from $K_{\ell 3}^0$ because hitherto we
have made no attempt to fine-tune isospin breaking effects.

\begin{figure}[h,t]
\centering{\
\epsfig{figure=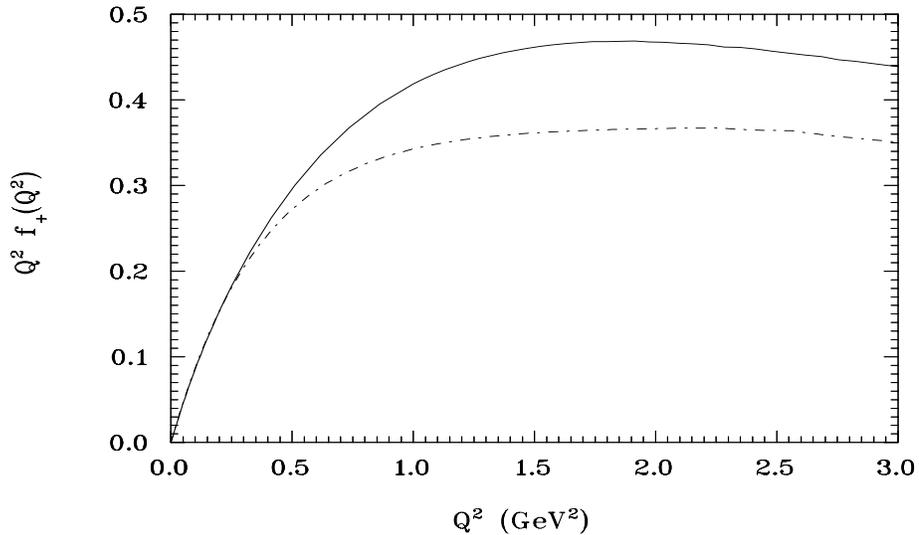,height=7cm,rheight=7cm,width=12cm,angle=90  }}
\caption{$-Q^2 f_+^K(Q^2)$: solid line; $-Q^2 f_+^\pi(Q^2)=Q^2 F_\pi(Q^2)$:
dot-dashed line.  The peak in $Q^2 f_+(Q^2)$, most pronounced for the kaon,
is a characteristic signal of quark-antiquark recombination into the meson
final state in exclusive processes.
\label{figb}}
\end{figure}
In Fig.~\ref{figb} we compare $f_+^K(Q^2)$ with $f_+^\pi(Q^2)$.  Since we
work in the isospin symmetric limit, $f_+^\pi(Q^2)=-F_\pi(Q^2)$. The results
are qualitatively similar to those obtained in Ref.~\cite{AB96}, although, as
in Ref.~\cite{BRT96}, our results for $f_+^{K \pi}(Q^2)$ are uniformly
smaller in magnitude.

{\bf 4. Summary and Conclusions}.\\ We have analysed the $K_{\ell 3}$ and
$\pi_{e 3}$ transition form factors.  An important new element of this study
is the development of an understanding of the form of the dressed
$s$-$u$-$W$-vertex, $V_\mu^{su}(k;Q)$, on the domain
$-(m_K+m_\pi)^2<Q^2<\infty$.  A simple heuristic Ansatz for $V_\mu^{su}$,
with one adjustable parameter, yields a uniformly good description of
$K_{\ell 3}$ observables, as illustrated in column two of Table~\ref{tableA}.

We identified a systematic discrepancy between the studies of
Refs.~\cite{SI92,AB96} and Refs.~\cite{LR84,GL85} arising because, in
Refs.~\cite{SI92,AB96}, nonanalytic contributions to $V_\mu^{su}$ in the
timelike region are overlooked.  These contributions are important on the
domain $0\lsim t \lsim (m_K^2-m_\pi^2)$ and allow for a correct description
of the scalar form factor, $f_0^K$, in this region; for example, without
these contributions it is not possible to reproduce the current-algebra
prediction for $f_0^K(t)$ at the Callan-Treiman point: $t=m_K^2-m_\pi^2$.
The difference between the results in columns one and two of
Table~\ref{tableA} is a measure of the relative importance of these
contributions to a given quantity.

Our results are consistent with the Ademollo-Gatto theorem, $f_+^2(0)\approx
1$.  They also support the expectation~\cite{NI75} that $f_-^K(0)$ is a
quantitative measure of the nonperturbative enhancement of the current-quark
mass difference $m_s-m_u$; i.e., it is a measure of the Euclidean
constituent-quark mass-ratio: $M^E_s/M^E_u$.

Figures~\ref{figa} and \ref{figc} illustrate that it is a good approximation
to consider $f_+(t)$ to be linear functions on the domain
$0<t<(m_K-m_\pi)^2$.  In common with other studies that employ quark-gluon
degrees-of-freedom, we find $\lambda_- \simeq \lambda_+$.  This suggests that
the assumption, employed in experimental analyses, that $f_-(t)$ is a
constant on this domain, should be reconsidered.

A strength of the DSE framework is that it allows the study of the
$t$-dependence of the transition form factors on the entire domain
$-\infty<t<(m_K-m_\pi)^2$.  In the isospin symmetric limit, $m_d=m_u$,
$f_+^\pi(t)= -F_\pi(t)$, the electromagnetic pion form factor and
$f_-^\pi(t)\equiv 0$.  One also has $f_+^{K^+}(t)=f_+^{K^0}(t)$, however,
$f_+^K(t)\neq -F_K(t)$, the electromagnetic kaon form factor.  There are
qualitative similarities; for example, $-t f_+^K(t)$ exhibits the peak
characteristic of quark-antiquark recombination into the meson final state in
this exclusive process, however, $|f_+^K(t)|$ falls-off less rapidly than
$|F_K(t)|$ at large-$(-t)$ and this is a measure of the difference between
the pion and kaon Bethe-Salpeter amplitudes; i.e., of nonperturbative effects
in QCD.

{\bf Acknowledgments}.~K.L.M. and Yu.K. are grateful for the hospitality of
the Physics Division at ANL during visits in which this study was conceived
and conducted.  The work of K.L.M was supported in part by the Natural
Sciences and Engineering Research Council of Canada; that of C.D.R. by the US
Department of Energy, Nuclear Physics Division, under contract number
W-31-109-ENG-38. The calculations described herein were carried out using the
resources of the National Energy Research Supercomputer Center.

%______________________________ References ______________________________

%___________________________________________________________________________
\end{document}